\documentstyle [twocolumn,epsf]{mn}
\newcommand{\mincir}{\raise
-2.truept\hbox{\rlap{\hbox{$\sim$}}\raise5.truept\hbox{$<$}\ }}
\newcommand{\magcir}{\raise
-2.truept\hbox{\rlap{\hbox{$\sim$}}\raise5.truept\hbox{$>$}\ }}
\newcommand{\minmag}{\raise
-2.truept\hbox{\rlap{\hbox{$<$}}\raise6.truept\hbox{$<$}\ }}
\newcommand{\be}{\begin{equation}}
\newcommand{\ee}{\end{equation}}
\newcommand{\ba}{\begin{eqnarray}}
\newcommand{\ea}{\end{esqnarray}}
\newcommand{\brr}{\begin{array}}
\newcommand{\err}{\end{array}}
\newcommand{\bc}{\begin{center}}
\newcommand{\ec}{\end{center}}

\title[Galaxy \& Cluster Bias]{Galaxy \& Cluster Biasing from Local Group
Dynamics}
\author[Plionis et al.]
{M. Plionis$^{1}$, S. Basilakos$^{1,2,3}$, M. Rowan-Robinson$^{3}$, 
S.J. Maddox$^{4}$, 
\newauthor
S.J. Oliver$^{3}$, O. Keeble$^{3}$, 
W. Saunders$^{5}$ \\
$^1$ National Observatory of Athens, Lofos Nimfon, Thesio, 18110 Athens,
Greece \\
$^2$ Physics Dept., University of Athens, Panepistimiopolis, Greece \\
$^3$ Imperial College of Science, Technology and Medicine, Blackell 
Laboratory, Prince Consort Road, London SW1 2EZ, UK \\
$^{4}$ Institute of Astronomy, Madingley Road, Cambridge, CB3 0EZ, UK\\
$^{5}$ Department of Astrophysics, Oxford University, Keble Road, OX1 3RH
}

\begin{document}

\maketitle

\begin{abstract}

Comparing the gravitational acceleration induced on the Local Group of galaxies
by different tracers of the underline density field 
we estimate, within the linear gravitational instability theory and the 
linear biasing ansatz, their relative bias factors.
Using optical SSRS2 galaxies, IRAS (PSCz) galaxies and Abell/ACO clusters,
we find $b_{O,I}\approx 1.21 \pm 0.06$ and $b_{C,I} \approx 4.3 \pm 0.8$, 
in agreement with other recent studies. Finally, there is an excellent 
one-to-one correspondence of the PSCz and Abell/ACO cluster dipole profiles, 
once the latter is rescaled by $b_{C,I}$, out to at least $\sim 150$ $h^{-1}$ 
Mpc.

\vspace{0.25cm}

\end{abstract}

\section{Introduction}

Different classes of extragalactic objects trace 
the underlying matter distribution differently. The realization of 
such a behaviour arose from the fact that the amplitude of the 
2-point correlation function of clusters of galaxies is significantly
higher than that of galaxies (cf. Bahcall \& Soneira 1983). This was 
suggested by Kaiser (1984) as a result of the clustering characteristics
of different height peaks in an underlying random Gaussian field.
A first order description of the effect is provided by linear 
biasing in which the extragalactic mass tracer fluctuation field is related 
to that of the underlying mass by: 
\be\label{eq:model}
\left(\delta \rho/\rho \right)_{\rm tracer} = b 
\left(\delta \rho/\rho \right)_{\rm mass}
\ee
with $b$ the linear bias factor. Even in this simplistic model the bias
cannot be measured directly and only theoretical considerations or numerical 
simulations can provide some clues regarding its value. However, the relative 
bias between different tracers can be measured and such attempts, using 
their clustering properties, have provided interesting, although somewhat 
conflicting, results.
Lahav, Nemiroff \& Piran (1990) comparing the angular correlation function of
different subsamples of the UGC, ESO and IRAS catalogues find an optical
to IR galaxy bias factor, $b_{O,I}$, ranging from 1 to 2 with preferred value
$b_{O,I} \sim 1.7$. Babul \& Postman (1990) using the
spatial correlation function of the CfA and IRAS galaxies find $b_{O,I} \simeq
1.2$, while comparing the QDOT correlation function (Saunders, 
Rowan-Robinson \& Lawrence 1992) with that of APM 
galaxies (Maddox et al. 1990) one finds $b_{O,I} \sim 1.4$. Similarly,
Oliver et al. (1996) comparing the clustering properties of
the APM-Stromolo survey of optical galaxies and an extended IRAS redshift 
survey found $b_{O,I} \sim 1.2 \pm 0.05$.
Strauss et al. (1992a) using the 1.936 Jy IRAS sample find that
the overdensity ratio between CfA and IRAS galaxies within a sphere 
centered on Virgo with the Local Group on the periphery gives $b_{O,I} 
\simeq 1.2$ while their correlation function analysis provides discrepant
results when comparing IRAS to CfA or SSRS optical galaxies (with $b_{O,I}\sim
2$ and 1, respectively). 
Recently, Willmer, daCosta \& Pellegrini (1998) using the SSRS2 sample
of optical galaxies and comparing with the clustering properties of the
1.2 Jy IRAS survey find $b_{O,I} \sim 1.2$ and $1.4$ ($\pm 0.07$) in redshift 
and real space respectively, while Seaborne et al (1999) performing a similar 
analysis between the PSC and Stromolo-APM redshift surveys find $b_{O,I} \sim 
1.3\pm 0.1$.

A different approach using the dynamics of the local group of galaxies, was 
proposed in Plionis (1995) and Kolokotronis et al. (1996). 
Traditionally, dynamical studies have been used in an attempt
to constrain the value of the cosmic density parameter, 
$\Omega_{\circ}$, by assuming linear theory and comparing observed 
galaxy or cluster peculiar velocities with estimated accelerations (cf. 
Strauss \& Willick 1995). However due to biasing
only the combination $\Omega_{\circ}^{0.6}/b$ can
be estimated. Such an analysis has been extensively applied to the Local 
Group of galaxies, since its peculiar velocity is accurately determined from 
the CMB temperature dipole (Kogut et al. 1996) and its gravitational 
acceleration can be measured from the dipole moment of the 
surrounding spatial distribution of different mass tracers. 
Within linear theory acceleration and peculiar velocity should be 
aligned and this indeed has been found to be the case using optical, 
IR galaxies, X-ray or optical cluster surveys and AGN's (cf. reviews of
Strauss \& Willick 1995, Dekel 1997 and references therein).
In the linear biasing framework the different mass tracers 
should therefore exibit similar dipole profiles differing only in 
their amplitudes, the ratio of which is a
measure of their relative bias. Therefore, one can estimate the 
relative bias factor between different mass tracers, because in the 
intercomparison of their velocity-acceleration relations the $\Omega_{\circ}$
parameter as well as the velocity cancels out.

In this study we use the recently completed PSCz IRAS galaxy survey 
(Saunders et al. 1999), the SSRS2 optical galaxy catalogue 
(DaCosta et al. 1998) and a subsample of the Abell/ACO cluster catalogue 
(as defined in Branchini \& Plionis 1996) to estimate their relative bias 
factors by comparing their dipole moments.

\section{Method}
Using linear perturbation theory one can relate the gravitational 
acceleration of an observer, induced by the
surrounding mass distribution, to her/his peculiar velocity:
\be
{\bf v(r)}=\frac{\Omega_{\circ}^{0.6}}{b} \frac{1}{4 \pi} \int 
\delta({\bf r})
\frac{\bf r}{|{\bf r}|^3} {\rm d}r = 
\frac{\Omega_{\circ}^{0.6}}{b} {\bf D}(r) 
\ee
The dipole moment, ${\bf D}$, is estimated by weighting
the unit directional vector pointing to the position of each tracer, with its
gravitational weight and summing over the tracer distribution; 
\be
{\bf D} =\frac{1}{4 \pi \langle n \rangle} 
\sum \frac{1}{\phi(r) \; r^{2}} \hat{{\bf r}}
\ee
with
\be\label{eq:sf}
\phi(r)=\frac{1}{\langle n \rangle} \int_{L_{{\rm min}}(r)}^{L_{{\rm max}}} 
\Phi(L)\,{\rm d}L
\ee
where $\Phi(L)$ is the luminosity function of the objects under study,
$L_{\rm {min}}(r)=4\pi r^{2} S_{\rm {lim}}$, with $S_{\rm {lim}}$
the flux limit of the sample
and $\langle n \rangle$ is the mean tracer number density, given by 
integrating the luminosity function over the whole luminosity range. 

\noindent
Using two different tracers, $i$ and $j$, of the underlying matter density
field to determine the Local Group acceleration one can write:
${\bf v(r)}= \Omega_{\circ}^{0.6} {\bf D}_{i}(r)/b_{i} = 
\Omega_{\circ}^{0.6} {\bf D}_{j}(r) /b_{j}$
and therefore we can obtain an estimate of their relative bias factor from:
\be \label{eq:bias1}
b_{ij}(r)=\frac{b_{i}}{b_{j}}(r) = \frac{{\bf D}_{i}}{{\bf D}_{j}}(r)
\ee
Since the dipole is a cumulative quantity and at each distance it depends 
on all previous shells, we cannot define an unbiassed $\chi^{2}$ statistic to
fit eq.\ref{eq:bias1}. Rather, we can obtain a crude estimate of the 
reliability 
of the resulting bias factor by estimating Pearson's correlation coefficient,
 $R_{i,j}$, between the two dipole profiles (see Kolokotronis et al. 1996); 
a value $R_{i,j} \simeq 1$ would indicate a perfect match of the two dipole
profiles and thus a very reliable estimate of their relative linear 
bias factor.

A statistically more reliable approach is to assume that the 
differential dipoles, estimated in equal volume shells, are independent 
of each other and then fit $b_{ij}$ according to:
\be \label{eq:bias2}
\chi^2 = \sum_{k=1}^{N_{bins}} \frac{({\bf D}_{i,k} - b_{ij} {\bf D}_{j,k} 
- {\cal C}_k)^{2}}{\sigma_{i,k}^2 + b_{ij}^{2} \sigma_{j,k}^2}
\ee
where $\cal{C}$ is the zero-point offset of the relation and $\sigma$ is
the corresponding shot-noise errors, estimated 
by using either of two methods; a Monte-Carlo
approach in which the angular coordinates of all tracers are
randomized while keeping their distance, and thus their selection function, 
unchanged or the analytic estimation of Strauss et al. (1992b); 
$\sigma^{2} \simeq \sum \phi^{-1}r^{-4} (\phi^{-1}+1)$.

\section{Data}

We use in our analysis three different catalogues of mass tracers;
\begin{itemize}
\item The recently completed IRAS flux-limited 
60-$\mu$m redshift survey (PSCz) which is described in
Saunders et al. (1999). It is based on the IRAS Point Source 
Catalogue and contains $\sim 15000$ galaxies with flux $>0.6$ Jy. 
The subsample we use, defined by $|b|\ge 10^{\circ}$ and limiting galaxy 
distance of 180 $h^{-1}$ Mpc, contains $\sim 10097$ galaxies and 
covers $\sim 82\%$ of the sky.
\item The SSRS2 catalogue of optical galaxies (DaCosta et al. 1998) which 
is magnitude limited to 
$m_{B}=15.5$ and contains 3573 galaxies in the South ($-40^{\circ}\le
\delta \le -2.5^{\circ}$, $b\le -40^{\circ}$) and 1939 galaxies in the North
($\delta \le 0^{\circ}$, $b\ge 35^{\circ}$), covering in total 13.5\% of the
sky.
\item A volume limited subsample of the Abell/ACO cluster catalogue,
with $|b|\ge 10^{\circ}$ and limited within 180 $h^{-1}$ Mpc
(see Branchini \& Plionis 1996). Our sample contains 197 clusters.
\end{itemize}

\subsection{Determining distances from redshifts}

All heliocentric redshifts are first transformed to the Local Group frame
using $c z \simeq c z_{\odot}+300 \sin(l)\cos(b)$. 
We then derive the distance of each tracer by using:
\be
r=\frac{2 c}{H_{\circ}} \left(1-(1+z-\delta z)^{-1/2} \right)
(1+z-\delta z)^{3/2}
\ee
where $H_{\circ} = 100 \; h$ Mpc and $\delta z$ is a non-linear term to correct
the redshifts for the tracer peculiar velocities:
\be \delta z =\frac{1}{c} ({\bf u}(r)-{\bf u}(0)) \cdot \hat{r} \ee
with ${\bf u}(0)$ the peculiar velocity of
the Local Group and ${\bf u}(r)$ the peculiar velocity of a galaxy or cluster
at position ${\bf r}$. 
Instead of using elaborate 3D reconstruction schemes (cf. Schmoldt et al 1999;
Branchini \& Plionis 1996; Branchini et al 1999; Rowan-Robinson et al. 1999) 
to estimate this term, 
we decided to use a rather simplistic velocity field model (see Basilakos 
\& Plionis 1998) to treat 
consistently all three data sets (a self-consistent 3D reconstruction of the
SSRS2 density field is in any case not possible due to the small area 
covered by the survey). 
Our simplistic velocity field model was found in Basilakos \& Plionis 
(1998) to be sufficient in order to recover the IRAS 1.2Jy and QDOT 
3-D dipole. We remind the reader the main assumptions of this model:

\noindent
{\bf (a)} The tracer peculiar velocities can be split 
in two vector components; that of a bulk flow and of a local non-linear term:
\be\label{eq:v-mod}
{\bf u}(r)={\bf V}_{bulk}(r)+{\bf u}_{\rm nl}(r)
\ee

\noindent
{\bf (b)} The first component dominates and thus that
\be {\bf u}(r) \cdot \hat{\bf r} \approx {\bf V}_{bulk}(r) \cdot \hat{\bf r}
\ee
We then use the observed bulk flow direction and profile,
as a function of distance, given by Dekel (1997) and combined with 
that of Branchini, Plionis \& Sciama (1996).
The zero-point, $V_{bulk}(0)$, and the direction of the bulk flow is
 estimated
applying eq.(\ref{eq:v-mod}) at $r=0$ and assuming, due to the ``coldness'' 
of the local velocity field (cf. Peebles 1988), that ${\bf u}_{\rm nl}(0)
 \simeq 
{\bf u}_{\rm inf} = 200$ km/sec (where $u_{\rm inf}$ is the LG infall velocity 
to the Virgo Supercluster).
\begin{figure}
\mbox{\epsfxsize=10.5cm \epsffile{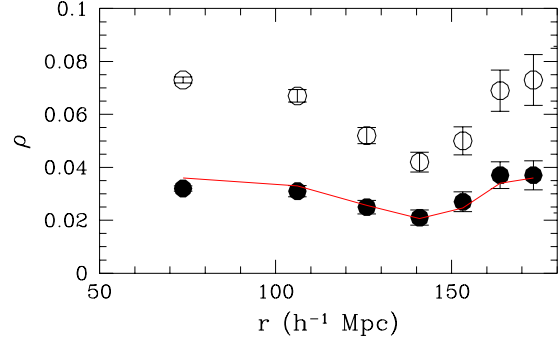}}
\caption{ Mean SSRS2 (open symbols) and PSCz (filled symbols)
galaxy space density estimated in equal volume shells with its Poissonian 
uncertainty. The continuous line is the SSRS2 density reduced by a factor 2. 
Based on 13.5\% of the sky (SSRS2 survey area).}
\end{figure}

\subsection{Galaxy densities}
To estimate the local acceleration field it is necessary to recover the true 
galaxy density field from the observed flux-limited samples. This is done
by weighting each galaxy by $\phi^{-1}(r)$, where $\phi(r)$, is defined in
eq.\ref{eq:sf}. 
For the PSCz sample we use the Saunders et al. (1990) luminosity function 
derived from the QDOT catalogue, with
$L_{min}=7.5\times 10^{7} \; h^{-2} L_{\odot}$ since lower luminosity galaxies
are not represented well in the available samples (cf. Rowan-Robinson et al. 
1990), and $L_{max}= 10^{13} \;  h^{-2} L_{\odot}$. 
For the SSRS2 sample we use the Schechter
luminosity function of Marzke et al. (1998) with $M_{max}=-22$ and
$M_{min}=-13.8$.

In figure 1 we present the mean density and its Poissonian 
uncertainty of PSCz and SSRS2 galaxies in their 
common area (that of the SSRS2 sample) and in equal volume shells (with
$\delta V \simeq 4.5 \times 10^6$ $h^{-3}$ Mpc$^{3}$). Their densities are 
extremely comparable, differing only by a constant factor 
($\langle \rho_{O}\rangle /\langle \rho_{I}\rangle=2.03 \pm 0.16$).

\section{Results}
\subsection{Optical to IR galaxy bias}
We first present the results of the intercomparison of the SSRS2 and PSCz
samples in their common sky area and for $r \ge 15$ $h^{-1}$ Mpc.
In figure 2a we show the amplitudes of the two dipoles as a function
of distance from the LG. The monotonic dipole increase reflects 
the fact that we are measuring only the component of the whole sky dipole 
which is due to the particular area covered by the sky restricted 
SSRS2 sample. It is apparent that the shapes of the two dipole 
amplitudes are extremely similar, giving correlation coefficient 
$R \simeq 0.97$. 

In figure 2b we present the direct dipole
ratio (eq.\ref{eq:bias1}) in the LG frame (open symbols), 
while as filled squares the results of the fit of eq.\ref{eq:bias2}, 
as a function of maximum distance used. No significant differences are found
when correcting distances for peculiar velocities. It is evident that the 
different estimates are consistent with each other, especially for $r>50$ 
$h^{-1}$ Mpc where the direct dipole ratio becomes flat.
It is essential, however, to verify whether such a good dipole-profile 
correlation could result solely due to the small solid angle used, ie., 
to investigate whether the survey geometry, coupled with the
galaxy selection function, dominates the dipole signal of both samples. To this
end we have generated 100 mock SSRS2 samples by reshuffling the galaxy angular 
coordinates while leaving their distances, and thus selection function, 
unchanged. If the reshuffled dipole profile resembles that of the SSRS2 
original one, then this would indicate the existence of the previously 
suggested bias. 
In figure 3 we present both (a) the dipole ratio between the original 
SSRS2 dipole and the reshuffled one together with its scatter and 
(b) the SSRS2 and PSCz dipole ratio with the latter rescaled by $b_{O,I}$. 
It is evident that the former ratio deviates 
significantly from one, an indication that our comparison is not dominated 
by the suspected bias.
\begin{figure}
\mbox{\epsfxsize=12cm \epsffile{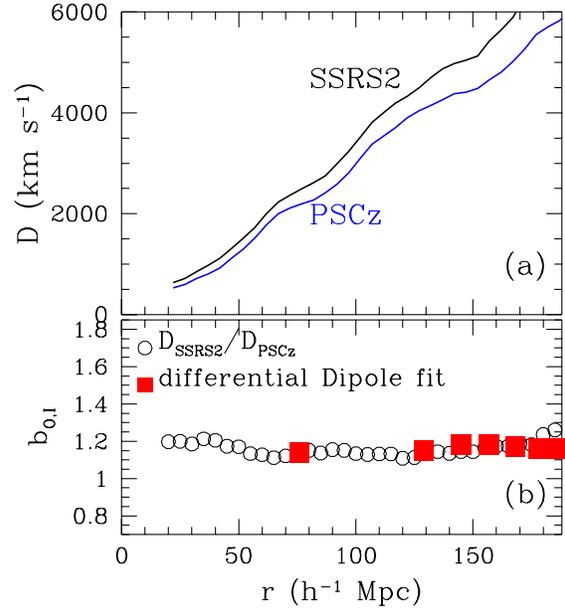}}
\caption{ (a) Dipole amplitude comparison for the SSRS2 and PSCz samples. 
(b) The estimated bias from comparing cumulative dipoles using 
eq.\ref{eq:bias1} (open circles) and differential dipoles using 
eq.\ref{eq:bias2} (filled squares).}
\end{figure}

\begin{figure}
\mbox{\epsfxsize=10.5cm \epsffile{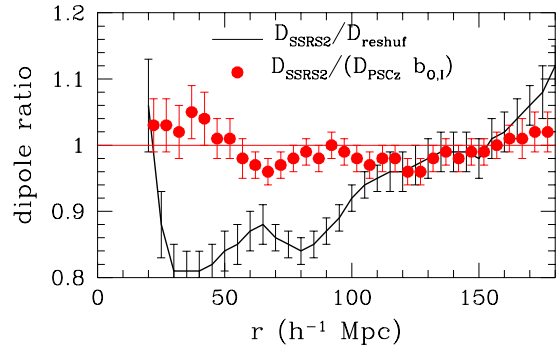}}
\caption{Cumulative Dipole Ratio test: Solid line represents the ratio of the 
original SSRS2 
and mean reshuffled dipoles; with its scatter estimated over 100 reshufflings 
of the galaxies angular coordinates. Solid points represent the
ratio of the SSRS2 and PSCz dipoles with the latter rescaled by 
$b_{O,I}\simeq 1.17$ (see text) while
the errors represents the propagation of the shot-noise dipoles.}
\end{figure}

Using the differential equal volume dipoles to fit eq.\ref{eq:bias2} for $10
\le r \le 185$ $h^{-1}$ Mpc, we find $b_{O,I} \simeq 1.24 \pm 0.04$,
with zero-point ${\cal C} \simeq -105 \pm 50$ km/sec and 
$\chi^{2}\approx 6.3$ for 6 degrees of freedom. 
The fit is performed using the {\small FITEXY} routine
of Press et al. (1992) and the uncertainties of the fitted parameters 
correspond to the $\Delta \chi^2 =1$ confidence region boundary. The small 
zero-point offset could be due to uncertainties in tracing the
very local contributions to the LG dipole. Indeed, integrating the the SSRS2 
and PSCz dipoles for $r> 15$ $h^{-1}$ Mpc (presented in figure 2) we find 
no zero-point offset 
${\cal C} \simeq -40 \pm 45$ km/sec but a slightly smaller bias factor
$b_{O,I} \simeq 1.17 \pm 0.04$, with $\chi^{2}\approx 7.5$ for 6 degrees of 
freedom. We derive a mean estimate of the bias factor and its 
uncertainty by varying both the inner and outer dipole integration limits in 
eq.\ref{eq:bias1} and by taking into account the slight differences  
of the results based on the differential dipole (eq.\ref{eq:bias1}).
The resulting optical to IR galaxy bias factor is:
$$b_{O,I} \simeq 1.21 \pm 0.06 \;\;.$$
Our result is in quite good agreement (within $1 \sigma$) with that of
Seaborne et al (1999), which is based on the PSCz and APM galaxy 
clustering properties on relatively smaller scales than those probed by 
our dipole analysis.

\subsection{Rich cluster to IR galaxy bias}
In the case of the cluster and PSCz samples we have a nearly 
full sky coverage, except at low-galactic latitudes. 
In order to recover the whole sky 
dipole we use a spherical harmonic approach to "fill" the unobserved part of 
the sky (cf. Lahav 1987; Plionis \& Valdarnini 1991). This approach has been 
found to provide compatible results with the cloning and interpolating method 
(see Branchini \& Plionis 1996 for such a comparison in the context of the 
cluster dipole).

The main drawback of comparing the cluster and galaxy dipoles arises from the 
fact that the Abell/ACO cluster distribution is incomplete in the local 
universe, since it does not include the Virgo cluster, an important 
contributor of the local velocity field (cf. Tully \& Shaya 1984). 
Therefore the direct 
comparison of the dipole amplitudes is hampered by this zero-point 
uncertainty.
We can attempt to correct for this problem by:

\noindent
$\bullet$ including the local Virgo contribution to the cluster dipole by 
assigning an appropriate Abell number count ($N_A$) weight to the 
Virgo cluster,

\noindent
$\bullet$ excluding from the PSCz dipole the very near contributions 
($\mincir 8 \; h^{-1}$ Mpc).

\noindent
A first attempt to derive the cluster to IR galaxy bias, using such a
procedure and the Abell/ACO and QDOT catalogues, was present in Plionis 
(1995). 
If clusters and galaxies do trace the same underline field, as indeed appears 
to be the case (cf. Branchini et al. 1999; Branchini, Zehavi, Plionis \& Dekel
1999), then we should be able to fit 
the two profiles, using eq.\ref{eq:bias2}, varying the Virgo cluster weight.
The appropriate value of $N_A$ is that for which the zero-point, $\cal{C}$, of
the fit vanishes and thus we will consider as our preferred bias 
parameter the corresponding value of $b_{C,I}$.
Statistically, this procedure does not provide a rigorous 
significance indication, due to the fact that the dipoles are 
cumulative quantities, but it provides a means of comparing quantitatively the
two dipole profiles. In order to test the robustness of the resulting bias 
parameter we fit the two dipole profiles as a function of distance. 

In figure 4a we present the resulting bias parameter versus the zero point, 
$\cal{C}$, for the Virgo cluster weights that provide a fit with 
$\cal{C}$ $\approx 0$. 
The different connected point arrays correspond to results based on the 
different $N_A$ weights while different points in 
each array correspond to different upper distance limits
used for the fit, which increase as indicated by the arrow. 
Taking into account that the zero-point uncertainty is $\delta{\cal C} \simeq 
130$ km/sec we conclude that the Virgo cluster weight for which 
${\cal C}\approx 0$ is $N_A = 24\pm 4$, confirming the notion that Virgo 
corresponds to a richness class $R=0$ Abell cluster. 
A consistency check, that we pass with success, is that 
the Virgocentric infall velocity that corresponds to this weight 
is $u_{\rm inf} \simeq 300 \pm 40$ km/sec, a value in agreement 
with most available determinations.

In figure 4b we 
present the fitted bias parameter as a function of upper distance limit 
(points) and the direct dipole ratio (eq.\ref{eq:bias1}) as broken lines for
the $N_A=24$ case. Both
seem consistent within their uncertainties, especially for distances 
$\sim 140 - 150 \; h^{-1}$ Mpc, which roughly 
corresponds to the apparent cluster dipole convergence depth.
\begin{figure}
\mbox{\epsfxsize=12cm \epsffile{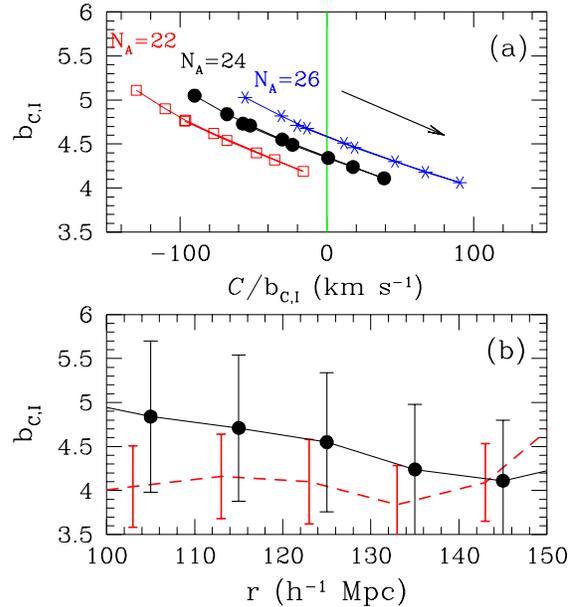}}
\caption{ (a) Fitted bias parameter between clusters and IRAS galaxies
versus zero-point, $\cal{C}$,  for different Virgo cluster weights. 
The different points within each 
connected array represent the results of the fit to different limiting
distance (increasing in the direction of the arrow). (b) The fitted bias
parameter (eq.\ref{eq:bias2}) as a function of limiting distance 
for $N_A=24$ (continuous line). The broken 
line is the corresponding direct dipole ratio (eq.\ref{eq:bias1}).}
\end{figure}
The main result regarding the bias parameter is:
$$ b_{C,I} \simeq 4.3 \pm 0.8 \;,$$
which interestingly is mostly independently of the Virgo cluster weights
(as can be seen in figure 4a), since such differences are 
absorbed in the value of $\cal{C}$. The uncertainty in $b_{C,I}$ reflects 
(a) variations due to different Virgo weights, (b) the variation between the 
eq.\ref{eq:bias1} and eq.\ref{eq:bias2} solutions and (c) the scatter around
these solutions (see figure 4b). It does not include, however, the cosmic 
variance which results from the use of only one observer.
Our results are in excellent agreement with those of Peacock \& Dodds (1994)
and Branchini et al. (1999b) based on completely different approaches.

\begin{figure}
\mbox{\epsfxsize=10.5cm \epsffile{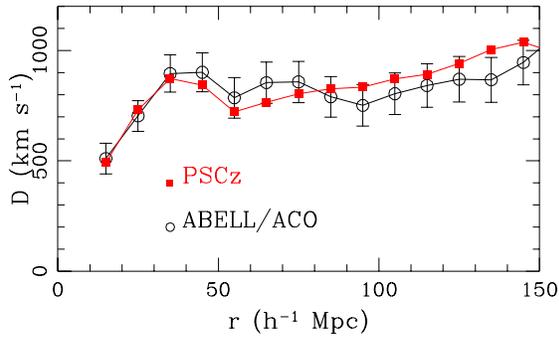}}
\caption{ Comparison between the PSCz and the ABELL/ACO dipoles, after 
scaling down the latter by a bias factor of 4.3.}
\end{figure}

In figure 5 we present a direct comparison of the PSCz and Abell/ACO cluster 
dipoles, out to 150 $h^{-1}$ Mpc, after having scaled down the latter 
by $b_{C,I}=4.3$. The errorbars
in the cluster dipole reflect mainly the uncertainty of the cluster
density between the Abell and ACO parts of the sample (see Plionis \& 
Kolokotronis 1998 and references therein). The two profiles 
are in excellent agreement, at least, up to $\sim 150 \; h^{-1}$ Mpc with 
correlation coefficient $R=0.86$. 
This is a further indication that the two density fields are consistent with 
each other out to these distances and supports the existence
of dipole contributions from large depths (see also Schmoldt et al. 1999; 
Basilakos \& Plionis 1998), suggestions which were first put forward
by Plionis (1988), on the basis of the Lick counts, by Rowan-Robinson et al. 
(1990) on the basis of the QDOT survey and by Scaramella et al. 
(1991) and Plionis \& Valdarnini (1991) on the basis of Abell/ACO clusters. 
A thorough investigation of the deep PSCz dipole (distances $>150 \; h^{-1}$ 
Mpc) will be presented in Rowan-Robinson et al. (2000).

\section{Conclusions}
We have used a novel approach, based on the Local Group dipole properties,
to estimate the relative bias parameter of different mass tracers.
We find that the optical to IR galaxy bias parameter is $b_{O,I} \simeq 1.21
\pm 0.06$, while the rich cluster to IR galaxy bias is
$ b_{C,I} \simeq 4.3 \pm 0.8$. Our results are in good agreement
with others based on different approaches. We find that the IR galaxy 
and rich cluster dipole profiles are extremely compatible 
once the latter is rescaled by $b_{C,I}$ out to at least $\sim 150 \; 
h^{-1}$ Mpc.

\section* {Acknowledgements}
S.B. thanks the Greek State Fellowship Foundation for financial support 
(Contract No 2669). We thank V. Kolokotronis for useful comments.

\end{document}